%% file: QCD-11-005_temp.tex
\pdfoutput=1

\documentclass[11pt,twoside,a4paper,cmspaper,final,collab]{cms-tdr}
\def\svnVersion{250002}\def\cmsCernNo{2013-194}\def\cmsCernDate{\today}\def\cmsMessage{Published in the Journal of High Energy Physics as \href{http://dx.doi.org/10.1007/JHEP06(2014)009}{\doi{10.1007/JHEP06(2014)009.}}}

\begin{document}\cmsNoteHeader{QCD-11-005}

\hyphenation{had-ron-i-za-tion}
\hyphenation{cal-or-i-me-ter}
\hyphenation{de-vices}

\RCS$Revision: 244408 $
\RCS$HeadURL: svn+ssh://svn.cern.ch/reps/tdr2/papers/QCD-11-005/trunk/QCD-11-005.tex $
\RCS$Id: QCD-11-005.tex 244408 2014-05-30 18:43:36Z hirosky $
\providecommand{\JETPHOX}{\textsc{jetphox}\xspace}
\providecommand{\pTg}{\ensuremath{p_{\mathrm{T}}^\gamma}\xspace}
\providecommand{\pTj}{\ensuremath{p_{\mathrm{T}}^{\text{jet}}}\xspace}
\providecommand{\etajet}{\ensuremath{\eta^{\text{jet}}}\xspace}
\providecommand{\etag}{\ensuremath{\eta^\gamma}\xspace}
\providecommand{\sieie}{\ensuremath{\sigma_{\eta \eta}}\xspace}
\providecommand{\isog}{\ensuremath{{\text {Iso}}^\gamma}\xspace}
\providecommand{\isoTRK}{\ensuremath{{\text {Iso}}_{\text{TRK}}}\xspace}
\providecommand{\isoECAL}{\ensuremath{{\text {Iso}}_{\text{ECAL}}}\xspace}
\providecommand{\isoHCAL}{\ensuremath{{\text {Iso}}_{\text{HCAL}}}\xspace}
\providecommand{\HoE}{\ensuremath{H/E}\xspace}
\providecommand{\tdXS}{{$\rd^3\sigma/(\rd\pTg\, \rd\etag\, \rd\etajet)$}\xspace}
\providecommand{\EtaBmax}{1.44\xspace}
\providecommand{\EtaEmin}{1.57\xspace}
\cmsNoteHeader{QCD-11-005} 
\title{Measurement of the triple-differential cross section for photon+jets production in proton-proton collisions at  $\sqrt{s}=7$\TeV}

\date{\today}

\abstract{
A measurement of the triple-differential cross section,
\tdXS, in photon+jets final states
using a data sample from proton-proton collisions at
$\sqrt{s}=7$\TeV is presented.  This sample corresponds to an integrated
luminosity of 2.14\fbinv collected by the CMS detector at the LHC.
Photons and jets are reconstructed within a pseudorapidity range of
$\abs{\eta}<2.5$, and are required to have transverse momenta in the range
$40 < \pTg < 300$\GeV and $\pTj>30$\GeV, respectively.
The measurements are compared to theoretical predictions from the
\SHERPA leading-order QCD Monte Carlo event generator and the
next-to-leading-order perturbative QCD calculation from \JETPHOX.
The predictions are found to be consistent with the data
over most of the examined kinematic region.
}

\hypersetup{%
pdfauthor={CMS Collaboration},%
pdftitle={Measurement of the triple-differential cross section for photon+jets production in proton-proton collisions at  sqrt(s)=7 TeV},%
pdfsubject={CMS},%
pdfkeywords={CMS, physics, QCD}}

\maketitle

\hyphenation{pa-ram-e-tri-za-tion}

Studies of events produced in proton-proton collisions containing a photon and
one or more jets in the final state provide a direct probe of
quantum chromodynamics (QCD)~\cite{DP1,DP2,DP3,DP4,DP5}.
The production cross sections, examined for various angular
configurations, are sensitive to contributions from the QCD hard-scattering
subprocesses and to parton distribution functions (PDFs) of the
proton~\cite{d'Enterria:2012yj,Carminati:2012mm}.
Measurements of these cross sections serve to constrain
PDF models and provide information for improving phenomenological
Monte Carlo models, as well as testing the applicability of
fixed-order perturbative calculations over a wide range of kinematic regions.
Photon+jets (direct photon) events are a major source of background
to standard model measurements, most notably
for the study of a light, neutral Higgs boson in the decay channel
$\PH\rightarrow\gamma\gamma$~\cite{:2012gu},
as well as beyond-the-standard-model searches for
signatures of extra dimensions~\cite{CMSDarkMatterLED}
and excited quarks~\cite{Excited}, among others.  Photon+jets
events can also be used to calibrate jet energies~\cite{jetresponse},
and to model the missing transverse energy distributions
attributed to the presence of noninteracting particles~\cite{metscale}.

This Letter presents a measurement of the triple-differential
cross section for photon+jets production using a data set collected
by the Compact Muon Solenoid (CMS) detector at the Large Hadron
Collider (LHC) from $\Pp\Pp$ collisions at $\sqrt{s}=7$ TeV.
The data correspond to an integrated luminosity of 2.14\fbinv.
This measurement spans a
transverse momentum range of $40 < \pTg < 300$\GeV and $\pTj>30$\GeV
for photons and jets, respectively.  It is performed in four regions of
pseudorapidity for the photon ($\abs{\etag} < 0.9$, $0.9 \leq \abs{\etag} < \EtaBmax$,
$1.56 \leq \abs{\etag} < 2.1$ and $2.1 \leq \abs{\etag} < 2.5$) and two
regions of pseudorapidity for the leading-transverse-momentum jet
($\abs{\etajet}<1.5$ and $1.5\leq \abs{\etajet}< 2.5$).
The dominant mechanisms for direct production of photons with large
transverse momentum are the Compton-like gluon scattering process
$\cPg\cPq\to\Pgg\cPq$
and the quark-antiquark annihilation process,
$\cPq\cPaq\to\Pgg\cPg$~\cite{PhysRevLett.73.388}.
The main background for these processes comes from the decay of neutral
hadrons, such as $\Pgpz$ and $\Pgh$ mesons, into nearly collinear pairs of
photons.  The expected background contribution from $\PW$+jets and
diphoton production is negligible.
This measurement spans an $x$ and $Q^2$ region of
$0.002\lesssim x \lesssim 0.4$ and
$1600\le Q^2\le 9\times 10^4\GeV^2$,
and extends the kinematic regions of photon \pt
covered by earlier
measurements~\cite{PhysRevD.57.67,Akesson:1986mr,Alitti1993174,Aktas:2004uv,Aaron:2007aa,Chekanov:2004wr,Chekanov:2006un,Abazov:2008er,ATLAS:2012ar,Aad:2013gaa,D0:2013lra}.
Measurements of the differential cross sections and ratios of the
differential cross sections for different angular configurations are
compared to theoretical predictions.

The CMS detector is a general-purpose,
hermetic detector providing large solid angle coverage for electromagnetic
and hadronic showers, charged particle tracks, and muons. The CMS
experiment uses a right-handed coordinate system, with the origin at the
nominal interaction point, with the $x$ axis pointing to the center of the LHC
ring, the $y$ axis pointing up (perpendicular to the plane of the LHC ring),
and the $z$ axis along the counterclockwise-beam direction. The polar angle
$\theta$ is measured from the positive $z$ axis and the azimuthal angle
$\phi$ in the $x$-$y$ plane. The pseudorapidity is defined by
$\eta=-\ln[\tan(\theta/2)]$. A full description of the CMS
detector
can be found in Ref.~\cite{CmsExperimentAtCernLHC}.  The subdetectors most
relevant to this analysis are the electromagnetic calorimeter (ECAL), the
hadron calorimeter (HCAL), and the silicon tracker.  These detectors are
located within a 3.8\unit{T} superconducting solenoid
of 6~m internal diameter.   The ECAL is a homogeneous calorimeter
composed of approximately 76\,000 lead tungstate crystals with segmentation
$\Delta\eta \times \Delta\phi =  0.0174\times 0.0174$
(where $\phi$ is measured in radians),
corresponding to a physical area of $22\times22\unit{mm}^{2}$ at the
front face of a crystal in the central barrel region ($\abs{\eta}<1.5$) and
$28.62\times28.62\unit{mm}^{2}$ in two endcap regions
($1.5<\abs{\eta}<3.0$).
The HCAL is a brass/scintillator sampling calorimeter with segmentation of
$\Delta\eta \times \Delta\phi = 0.087\times 0.087$
in the central region ($\abs{\eta} <$ 1.74) and
$\Delta\eta \times \Delta\phi = 0.09\times0.174$ to $0.35\times0.174$ for
forward pseudorapidity ($1.74 < \abs{\eta} < 3.0$). The silicon tracking
system, located between the LHC beam pipe and the ECAL, consists of
pixel and strip detector elements covering the pseudorapidity
range $\abs{\eta}<2.5$.  In the forward region a  preshower detector,
consisting of two planes of silicon sensors interleaved with 3
radiation lengths of lead, is located in front of the ECAL, covering the
region $1.65<\abs{\eta}<2.6$.

Events selected for this analysis are recorded using a
two-level trigger system.  A level-1
trigger requires a cluster of energy deposited in the ECAL
with transverse energy $\ET > 20$\GeV.  The CMS high-level trigger (HLT)
applies a more sophisticated energy clustering algorithm to
events passing the level-1 threshold and further requires \ET
trigger thresholds from 30 to 135\GeV.
These thresholds are raised with increased
instantaneous luminosity to prevent saturation of the readout.
In addition to these trigger requirements, an offline requirement is imposed
to ensure that events have at least one well reconstructed primary vertex
within 24\cm in $z$ of the nominal center of the detector.

Photons deposit most of their energy through electromagnetic showers
in the ECAL. They are reconstructed by clustering energy deposits
in neighboring crystals according to criteria that are optimized for
different regions of pseudorapidity.  Each clustering algorithm begins
from a seed crystal with large transverse energy.  In the barrel region,
clusters are formed by summing energies across 5 (35) crystals in the
$\eta$ ($\phi$)
direction.  Clusters in the endcap are formed by combining contiguous
$5\times5$ arrays of crystals and including the corresponding energy in
the preshower detector.  The full details of these algorithms can
be found in Ref.~\cite{EGM-10-006}.
We apply the same selection criteria used in the measurement of the
inclusive photon cross section~\cite{PhysRevLett.106.082001} and provide
a summary here.  A photon reaching the ECAL without undergoing conversion
to an $\Pep\Pem$ pair deposits most of its energy in a
$3\times3$ crystal matrix.  Only a very small fraction of the energy
from the resulting shower leaks into the HCAL,
hence the ratio of the energy of the photon candidate in the HCAL to the
energy in the ECAL, \HoE, within a cone of radius
$R=\sqrt{(\Delta\eta)^{2}+(\Delta\phi)^{2}}=0.15$ around the seed crystal
can be used to separate photon showers from electromagnetic components
of hadron-initiated showers.  For this analysis, a requirement
of $\HoE<5\%$ is applied to the photon candidates. To reject
electrons, we require that there be no hits in the first two inner
layers of the silicon pixel detector that are consistent with an electron
track matching the location and energy of the photon candidate in the
calorimeter (pixel detector veto).  To further improve the purity of the photon
candidate sample, an additional requirement is applied based on the
second moment of the electromagnetic shower in $\eta$,
calculated using a $5\times 5$ matrix of crystals around the
highest energy crystal in the cluster,

\begin{equation}
    \label{eq:pho_showcov}
    \sigma^2_{\eta \eta} = \frac
    { \sum \left( \eta_i - \bar\eta \right )^2  w_i
    } {\sum w_i},
\end{equation}

\noindent where the sum runs over all elements of the $5\times 5$ matrix,
and $\eta_i=0.0174\hat\eta_i$, with $\hat\eta_i$ denoting the $\eta$ index of the
$i$th crystal; the individual weights $w_i$ are given by
$w_i = \max \left( 0, 4.7 + \ln(E_i / E_{5\times5}) \right )$ and $E_i$ is the
energy of the $i$th crystal; $\bar\eta=\sum\eta_iw_i/\sum w_i$
is the energy-weighted average pseudorapidity.
The requirement $\sieie<0.01\,(0.028)$ in the barrel (endcaps) further
suppresses background from neutral mesons ($\Pgpz$, $\Pgh$, etc.)
that may satisfy the isolation
requirements described below as a result of fluctuations in the fragmentation
of partons. The combined \HoE and shower shape requirements along with
the pixel detector veto comprise the photon identification criteria.
If multiple photons
are reconstructed within the fiducial range of this analysis, only
the photon with highest \pTg, leading photon, is considered.

Jets are reconstructed using the anti-\kt~\cite{AntiKtAlgo} clustering
algorithm with distance parameter of 0.5.  Inputs for the
jet clustering are defined by the particle-flow~\cite{ParticleFlowAlgo}
algorithm, which is a full-event reconstruction technique that aims
to reconstruct and identify all stable particles produced in an event through
the combination of information from all subdetectors. Jets with
 $\pt> 30\GeV$ are selected for this analysis, and
are required to pass data quality requirements designed to remove
spurious jets resulting from noise~\cite{QCD-10-011}.
Inefficiencies due to these criteria are negligible.
Since energetic photons are also
reconstructed as jets by the anti-\kt algorithm, any jet that overlaps
with the leading photon within a cone of $R<0.5$ is removed from
consideration.

Even after the photon identification criteria are applied, a significant
background remains, mostly from neutral
mesons that decay to photons that overlap in the ECAL.
Templates constructed from signal and background distributions
are fitted to data to determine the purity of the selected photon sample.
The method exploits the distribution of energy in the vicinity of the
photon using the variable $\isog=\isoTRK+\isoECAL+\isoHCAL$,
where \isoTRK is the sum of the \pt of tracks consistent with the
reconstructed vertex in a hollow cone, $0.04<R<0.40$,
centered around the candidate photon momentum vector extending from
the primary vertex
to the ECAL cluster.  Similarly, \isoECAL is the transverse energy
deposited in the ECAL in $0.06<R<0.40$, and \isoHCAL is
the transverse energy deposited in the HCAL in $0.15<R<0.40$.
For the $\text{Iso}_{\text{TRK}}$ $(\text{Iso}_{\text{ECAL}})$
distributions, we do not include energy in a rectangular strip of
$\Delta \eta \times \Delta \phi=0.015$ $(0.040) \times 0.040$ to exclude
energy associated with the photon in case of conversion~\cite{EGM-11-001}.
The method takes advantage of differences in the \isog distributions between signal
and background. The main contribution to \isog for genuine photons
comes from the underlying event and multiple pp interactions
in the same bunch crossing (pile-up collisions). The average number of pile-up
collisions for data used in this analysis is ${\sim}6$.
In contrast,
\isog for misidentified photons includes additional contributions of energy
from jet fragmentation. Hence, the \isog distribution for the background
tends to be broader than for signal.

\begin{figure*}[hbtp]
\centering
 \includegraphics[width=0.5\textwidth]{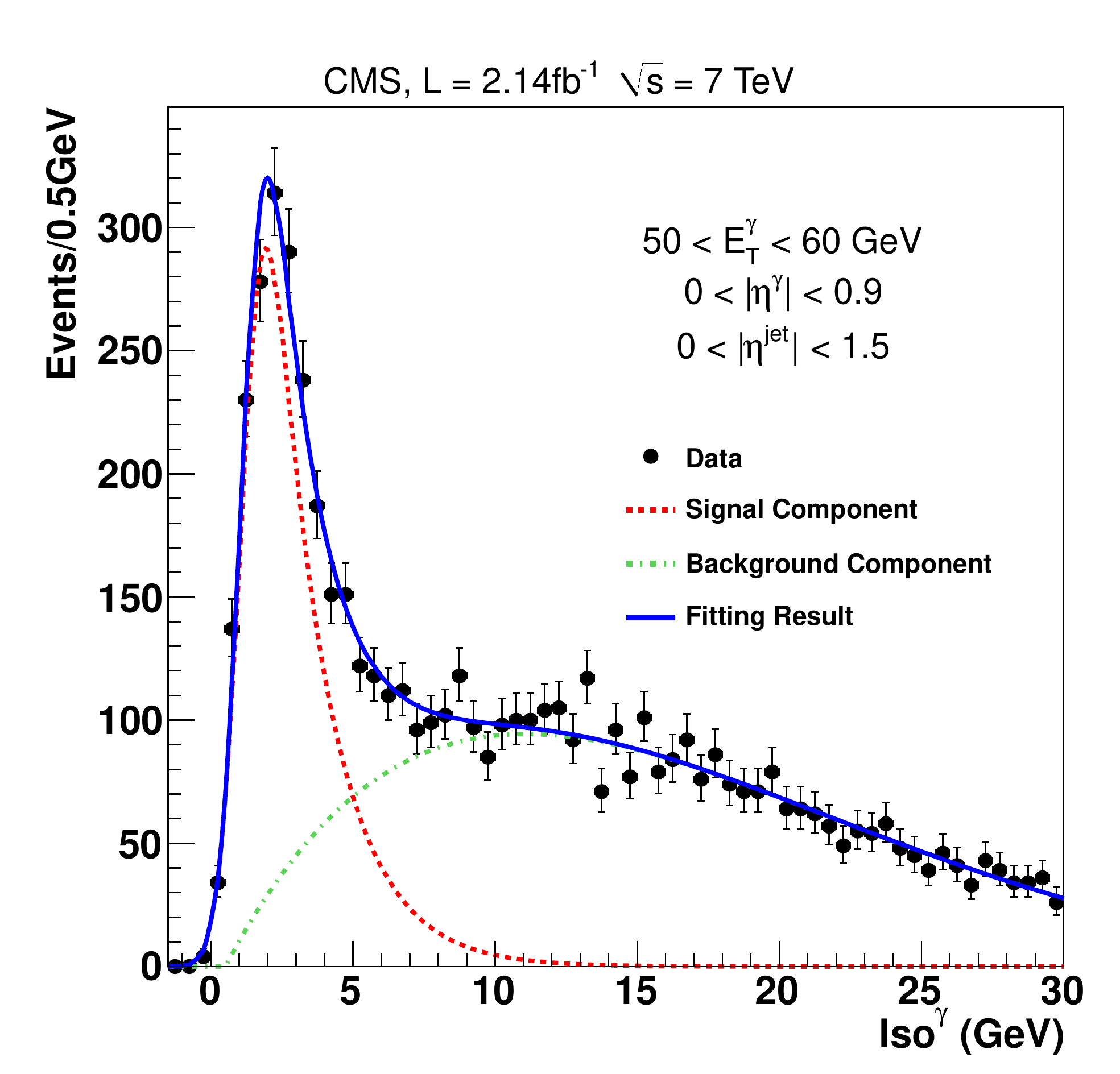}
 \caption{Example of a fit to the \isog\ distribution using signal and
background templates. }
\label{fig:isolationtemplatefit}
\end{figure*}

The signal template is modeled using Monte Carlo (MC)
events generated with \PYTHIA 6.424~\cite{pythia} and parameterized by
the convolution of an exponential function with a Gaussian,

\begin{equation}
    \label{eq:signal_model}
    S(x)= C_S\,
    \re^{\alpha x} \otimes \text{Gaussian}(x,\mu,\sigma),
\end{equation}

\noindent where $x=\isog$, $(\mu,\sigma)=\vec p$ and $\alpha$ describe
the peak and tail of the signal template, respectively, and $C_S$ normalizes
the distribution to unit area.
The background template is obtained from data using a background-enriched
sample collected from a sideband region, obtained by inverting the
shower shape selection requirement and requiring
$\sieie >0.011\,(0.030)$ in the barrel (endcap) regions.
The background distribution is parameterized
using an inverse ARGUS function ~\cite{Albrecht1990278},
\begin{equation}
\label{eq:bkg_model}
B(x)=
 \left\{
  \begin{array}{cl}
   C_B\left[ 1-\re^{z(x-q_1)} \right]\cdot[1-q_2(x-q_1)]^{q_3} & ;~x\geq q_1 \\
          0 & ;~x < q_1,
  \end{array}
 \right.
\end{equation}
 where $x=\isog$, $z$ describes the shape of the background template in the
signal-dominated region, $q_1$ ($q_2,q_3)$ describe
the starting point of the background template (or
its shape in the background-dominated region), and $C_B$ normalizes
the distribution to unit area.

The signal purity is determined by fitting the signal and background
template functional forms to data, $N_{S}\cdot\isog_{S} + N_{B}\cdot\isog_{B} $,
and minimizing an extended $\chi^{2}$ defined as
\begin{equation}
\chi^{2} =
\sum^n_{i=1}\left (\frac{N_i - (N_S S_i(\vec{p},\alpha) + N_B B_i(z,\vec{q}))}{\sigma_{N_i}}\right )^2
+ \left (\frac{(z - z_{\text{central}})}{\sigma_{z}}\right )^{2},
\label{eq:chi}
\end{equation}
where $N_{S}$ and $N_{B}$ are the numbers of signal and background events,
$n$ is the number of bins in the templates, $N_i$ the observed number
of events for the $i$th bin with uncertainty $\sigma_{N_i}$,  $S_i$ and
$B_i$ are the per-bin integrals of the corresponding signal and background
templates, and $z_{\text{central}}$ ($\sigma_z$) is the value (uncertainty) of the parameter $z$
determined by the fitting of the background template.
The parameters can be categorized into those that most directly model the
signal-dominated ($\mu$, $\sigma$, $z$, and $q_{1}$) and background-dominated
($\alpha$, $q_{2}$, and $q_{3}$) regions.  The parameter that describes
the peak in the signal template is allowed to vary in the fit to correct for
differences between data and MC in the region of low isolation energy.
This procedure is validated with data using a photon sample collected from
$\cPZ\to\Pgmp\Pgmm\Pgg$ events.
The parameter that describes the tail of the signal template in the
high isolation energy region is shifted by 5\% to account for
differences observed between data and MC simulation,
and to estimate the uncertainty from the contributions of nonprompt
photons, which originate from jet fragmentation.  In the low \isog\
region, the background distribution is constrained by the sideband data,
allowing the parameter $z$ to vary based on the value $z_{\text{central}}$
with an uncertainty $\sigma_z$. An example of the resulting templates is
shown in Fig.~\ref{fig:isolationtemplatefit}.
The purity is determined
independently in bins of $\gamma$ and jet pseudorapidity and as a function of
\pTg.

\begin{figure*}[thbp]
\centering
\includegraphics[width=0.4\textwidth]{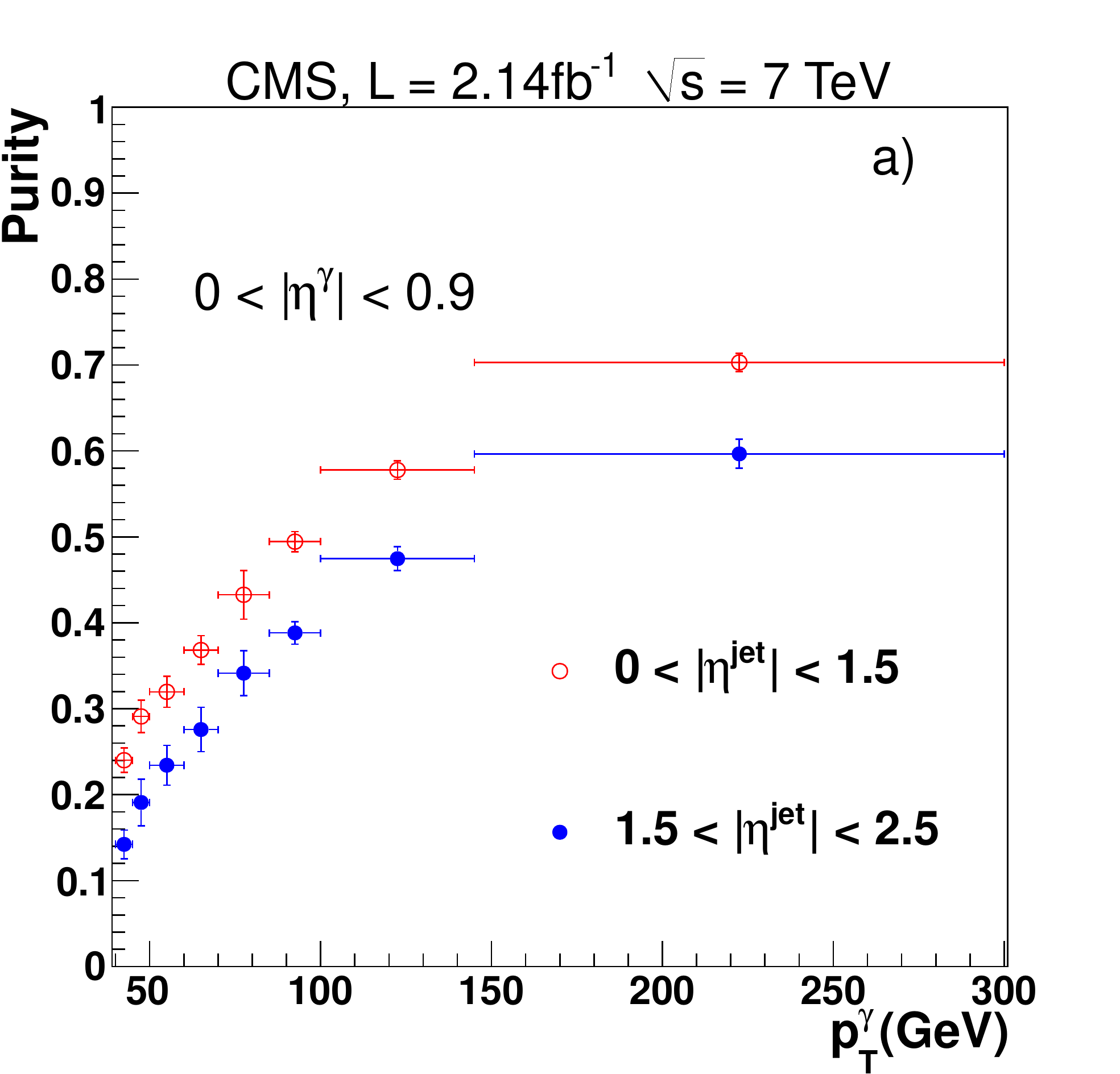}
\includegraphics[width=0.4\textwidth]{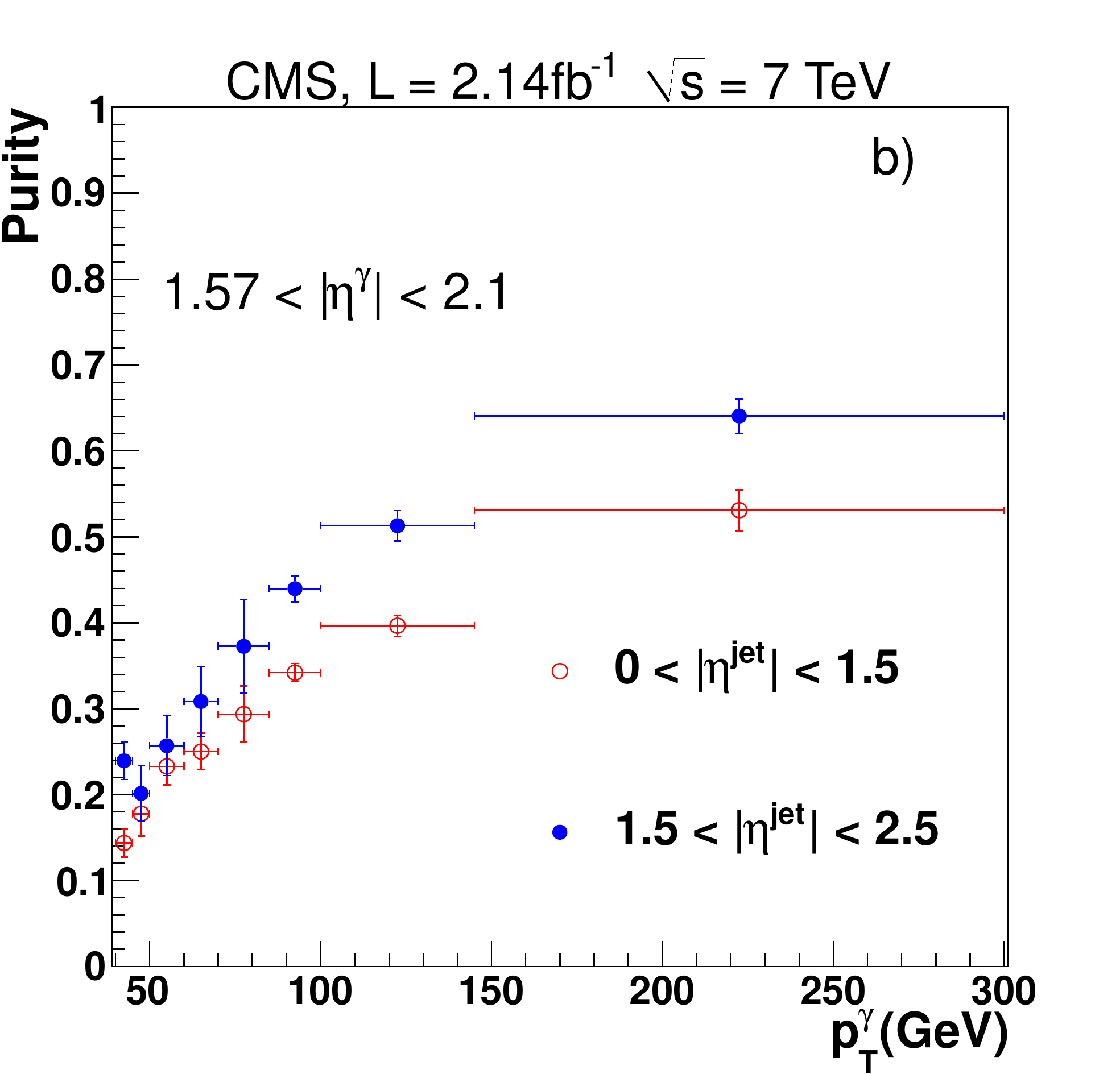}
\caption{Examples of signal purity as a function of \pTg for
(a) photons in the barrel and (b) photons in the endcap.
In each figure the open (filled) circles correspond to the events
with leading jet located in the barrel (endcap). The error bars
represent the total statistical and systematic uncertainty in the purity
measurement. }
\label{fig:purity_final_plots}
\end{figure*}

The signal purity is defined as the ratio of prompt photons to the total
number of selected photons. This is shown as a function of \pTg in
Fig.~\ref{fig:purity_final_plots} for two ranges of $\eta^{\gamma}$;
it increases with the transverse momentum of the photons.
The variation of the measured photon purity across kinematic
regions is consistent with expectations for signal and background,
which correspond to different admixtures of initial partonic states.
The main contribution to the systematic uncertainty in the photon signal
purity is due to the modeling the shape of the background
template, which is dominated by statistical uncertainty in the sideband samples.
This uncertainty is evaluated by performing pseudo-experiments based on
simulated QCD samples to examine variations in the measurement of the purity
due to statistical fluctuations in the template models.
The upper limit of $\pt^\gamma<300$\GeV in this analysis is
determined by the availability of data that allows us to independently
model background templates for each bin in the triple differential cross
section.
We also consider a contribution to the systematic uncertainty
related to the modeling
of the signal template.  The systematic uncertainty is evaluated independently
for each bin and increases with decreasing photon transverse momentum from
1\% to 30\%.

\begin{figure*}[hbtp]
\centering
 \includegraphics[width=0.6\textwidth]{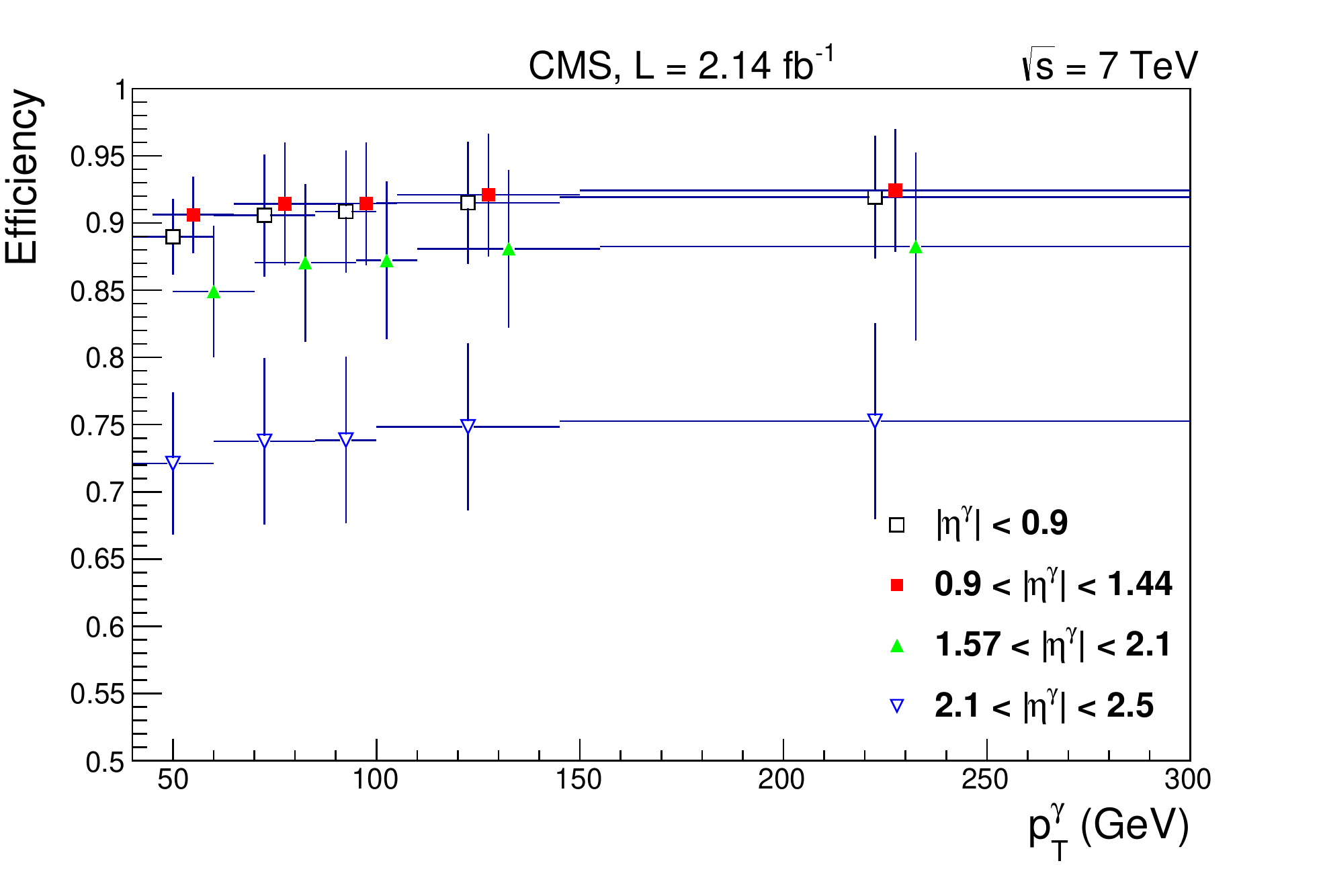}
 \caption{Total efficiency for photon selection as a function of photon
transverse momentum ($\pTg$) in four different photon pseudorapidity
($\etag$) ranges.
The error bars include both statistical and systematic uncertainties
and are dominated by the latter.
For clarity, points corresponding to the second and third
\etag bins are shifted to the right by 5 and 10\GeV, respectively.
}
\label{fig:totalefficiency}
\end{figure*}

The selection efficiency for photons can be factorized into four terms,
which are measured independently:
$\epsilon_{\text{total}} = \epsilon_{\text{trigger}}\cdot \epsilon_{\text{RECO}}
\cdot \epsilon_{\text{ID}}\cdot \epsilon_{\mathrm{PMV}}$.
The first factor, $\epsilon_{\text{trigger}}$, is the trigger selection
efficiency, and is measured in data using electrons from the decay of
$\Z$ bosons following a `tag-and-probe' method~\cite{CMS_NOTE_EGM-07-001}.
The tag electron is required to match an object reconstructed as an
HLT electron, while the probe requirement is relaxed to pass the offline photon
selection requirements and a photon HLT path.  This efficiency factor is
found to be consistent with 100\% within its systematic uncertainty.  The
reconstruction efficiency, $\epsilon_{\text{RECO}}$, is measured using simulated
events in a photon+jets sample generated with \PYTHIA. The same sample
is used to determine $\epsilon_{\text{ID}}$, the efficiency of
the photon identification criteria apart from the pixel detector veto.
The systematic uncertainty in the photon identification efficiency is
determined from the differences between MC simulation and data by applying
the nominal photon selection criteria to electrons in a Z-boson-enriched
data sample.
The photon pixel veto efficiency,
$\epsilon_{\mathrm{PMV}}$, is estimated from data by employing the
tag-and-probe technique with final-state-radiation photons in
$\Z\to\Pgmp\Pgmm\Pgg$ events and is independent of \pt.
The total photon efficiency as a
function of photon transverse momentum in the four photon pseudorapidity
ranges is shown in Fig.~\ref{fig:totalefficiency}.  The variation of total
efficiency values in the photon pseudorapidity regions is mainly caused
by the pixel veto efficiency contribution.

\begin{figure*}[!ht]
\centering
\includegraphics[width=0.6\textwidth]{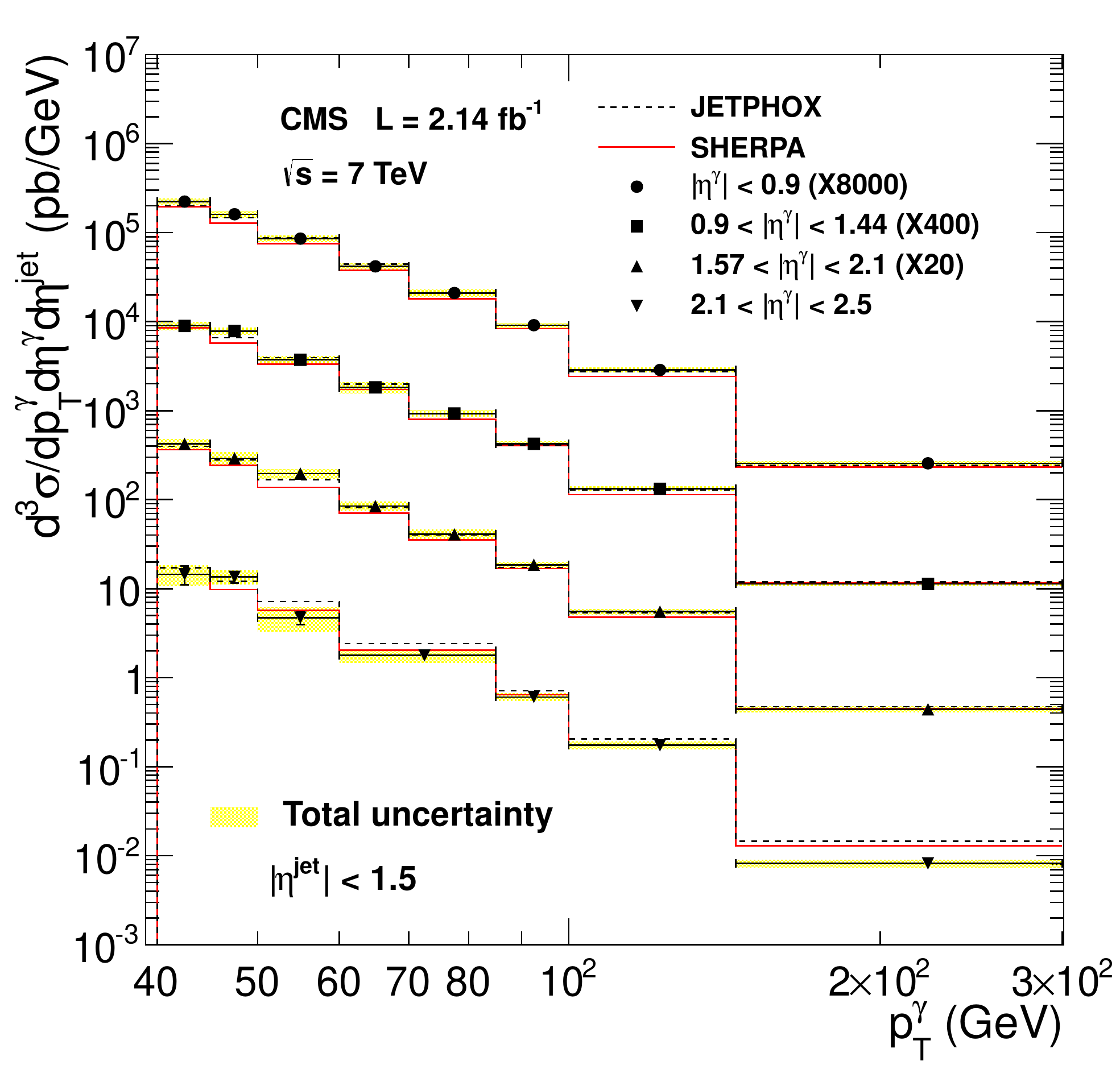}
\caption{Differential cross sections for $\abs{\etajet}<1.5$. The measured cross
sections (markers) in four different ranges of $\etag$ are compared with the
\SHERPA tree-level MC (solid line) and the NLO perturbative QCD calculation from
\JETPHOX (dashed line). The cross sections for the most central photons
are scaled by factors of 20 to 8000 for better visibility.
Error bars are statistical uncertainties and the shaded
bands correspond to the total experimental uncertainties.}
\label{fig:crossSection1}
\end{figure*}

\begin{figure*}[!ht]
\centering
 \includegraphics[width=0.6\textwidth]{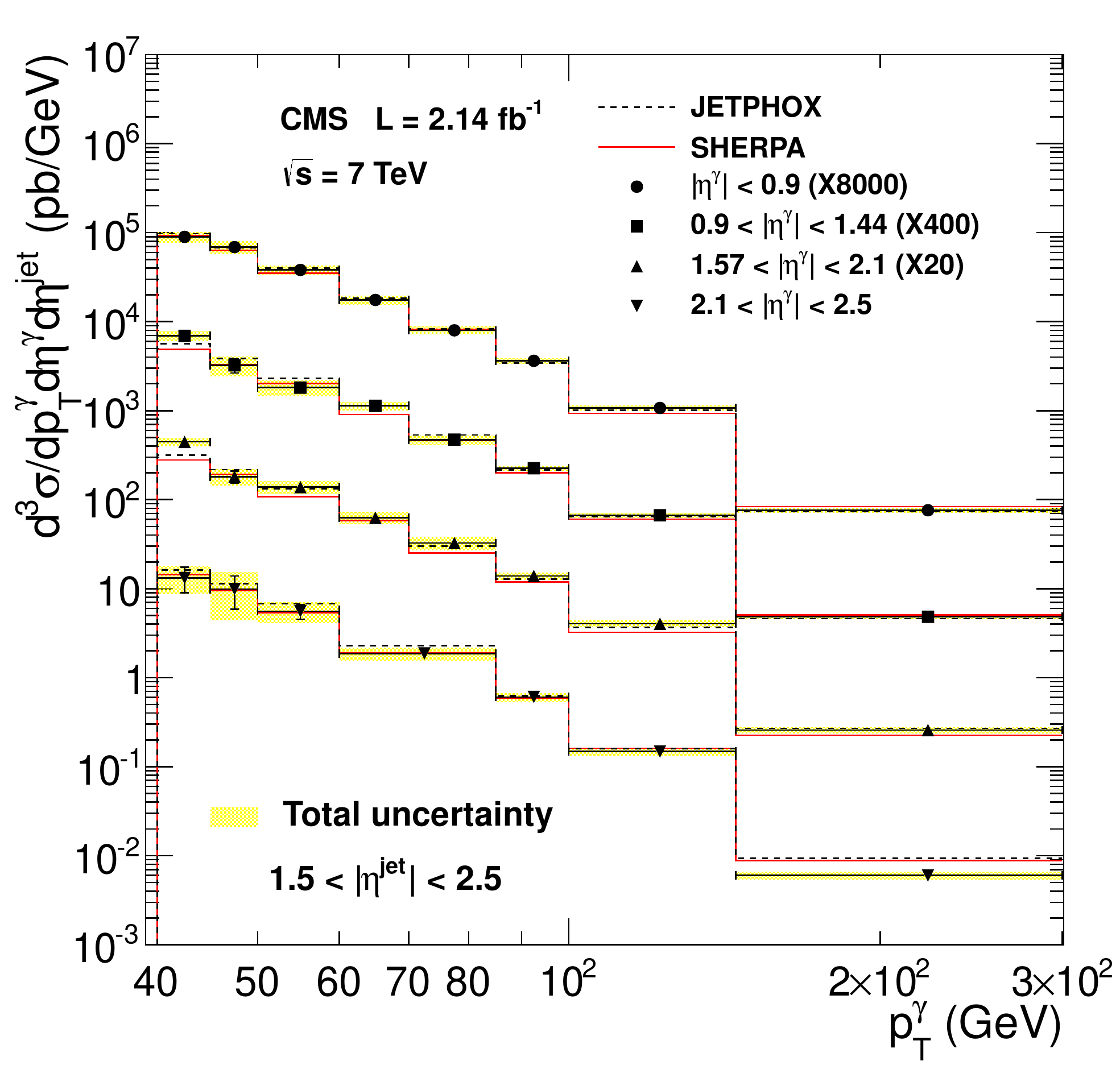}
 \caption{Differential cross sections for $1.5<\abs{\etajet}<2.5$. The measured
cross sections (markers) in four different ranges of $\eta^{\gamma}$ are compared
with the \SHERPA tree-level MC (solid line) and the NLO perturbative QCD
calculation from \JETPHOX (dashed line).
The cross sections for the most central photons
are scaled by factors of 20 to 8000 for better visibility.
Error bars are statistical
uncertainties and the shaded bands are the total experimental uncertainties.}
\label{fig:crossSection2}
\end{figure*}

Figures~\ref{fig:crossSection1}~and~\ref{fig:crossSection2} show the
measurement of the triple-differential cross section
\tdXS for $\abs{\etajet}<1.5$ and
$1.5<\abs{\etajet}<2.5$.  The measurements are corrected for
detector acceptance, efficiency and resolution
by unfolding the spectra using an iterative
method~\cite{ref:D'Agostini:1994zf}. The cross section
is calculated using

\begin{equation}
\label{XScalc}
\frac{\rd^3\sigma}{\rd{}\pt^{\gamma}\rd\eta^{\gamma}\rd\eta^{\text{jet}}}=
\frac{1}{\Delta \pt^{\gamma}\cdot\Delta\eta^{\gamma}\cdot\Delta\eta^{\text{jet}}}
\frac{N_{\text{signal}}^{\gamma}\cdot U}{L\cdot \epsilon},
\end{equation}

\noindent where $N_{\text{signal}}^{\gamma}$ is the number of photon candidates 
corrected for signal purity in bins of
$\Delta \pt^{\gamma}, \Delta \eta^{\gamma}$, and $\Delta \eta^{\text{jet}}$ with
integrated luminosity $L$; $U$ and
$\epsilon$ are the unfolding and efficiency corrections, respectively.

The systematic uncertainty due to the unfolding procedure is estimated
by varying the parameterization of signal model
and photon energy resolution according to differences between the data and MC
distributions.  The effect of uncertainties in the photon
energy scale varies from 1.1\%--1.5\%~(2.2\%--3\%) for
measurements with photons in the barrel (endcap),while those due to the jet
energy scale are negligible.
The contributions to the systematic uncertainty in the differential
cross section from the determination of photon reconstruction efficiency,
unfolding, photon energy scale and the photon purity determination
are given in Table~\ref{tab:crossSection_systematics}.  The table also
shows the total systematic uncertainty obtained by adding all the contributions
in quadrature. At low $\pTg$ the systematic uncertainty is dominated by
the purity determination.  This is also the region where the uncertainty
is the highest.  At high $\pTg$ the most significant contribution usually
comes from the determination of the reconstruction efficiency.

\begin{table}[!ht]
    \centering
 \caption{Contributions to the relative systematic uncertainty (in percent) in
the cross section measurement from efficiency, unfolding, photon energy scale,
and purity calculations.
The total systematic uncertainty is obtained by adding all the contributions in
quadrature, a 2.2\% uncertainty due to the integrated luminosity measurement
is not included.
The numbers in the table represent the ranges of uncertainties
obtained in different $\etag$ and $\etajet$ bins.}
    \label{tab:crossSection_systematics}
    {\small
    \begin{tabular}{|c|cllll|}
    \hline
 \multicolumn{6}{|c|}{$\abs{\etag} < \EtaBmax$}\\
    \hline
$\pTg$ (\GeVns{})  &  efficiency (\%)  &  unfolding (\%)  & photon energy (\%)   &  purity (\%)    &  total (\%) \\
\hline
40--45        &     2.5           &    2.1           &    1.1               &  4.9--9.3     &  6.0--10.0 \\
45--50        &     1.2           &    2.5           &    1.2               &  4.9--17      &  5.6--17 \\
50--60        &     4.5           &    2.6           &    1.4               &  4.2--13      &  6.8--14 \\
60--70        &     4.5           &    2.4           &    1.5               &  3.7--11      &  6.5--13 \\
70--85        &     4.5           &    1.2           &    1.5               &  4.6--5.7     &  6.7--7.5 \\
85--100       &     4.5           &    1.4           &    1.5               &  2.2--3.1     &  5.4--5.8 \\
100--145      &     4.5           &    1.4           &    1.5               &  1.8--2.5     &  5.2--5.6 \\
145--300      &     4.5           &    1.2           &    1.5               &  1.4--2.6     &  5.1--5.5 \\
\hline
 \multicolumn{6}{|c|}{$\EtaEmin < \abs{\etag} < 2.5$}\\
    \hline
$\pTg$ (\GeVns{})  &  efficiency (\%) &  unfolding (\%) & photon energy (\%) &  purity (\%) &  total (\%)\\
  \hline
40--45        &     3.0          &     2.1         &      2.2           &  6.9--9.9  &  8.1--11 \\
45--50        &     3.5          &     2.5         &      2.4           &  8.6--38   &  9.9--38 \\
50--60        &     5.0          &     2.6         &      2.7           &  7.2--25   &  9.5--25 \\
60--70        &     5.0          &     2.4         &      3.0           &  7.0--12   &  9.4--14 \\
70--85        &     5.0          &     1.2--5.0  &      3.0           &  10--13    &  12--15  \\
85--100       &     5.0          &     1.4--5.0  &      3.0           &  2.8--4.6  &  6.6--8.6 \\
100--145      &     5.0          &     1.4--4.0  &      3.0           &  2.8--6.3  &  6.6--8.7 \\
145--300      &     5.0          &     1.2--2.1  &      3.0           &  2.9--5.1  &  6.8--7.9 \\
\hline
\end{tabular}
    }
\end{table}

\begin{figure*}[!ht]
\begin{center}
\includegraphics[width=0.7\textwidth]{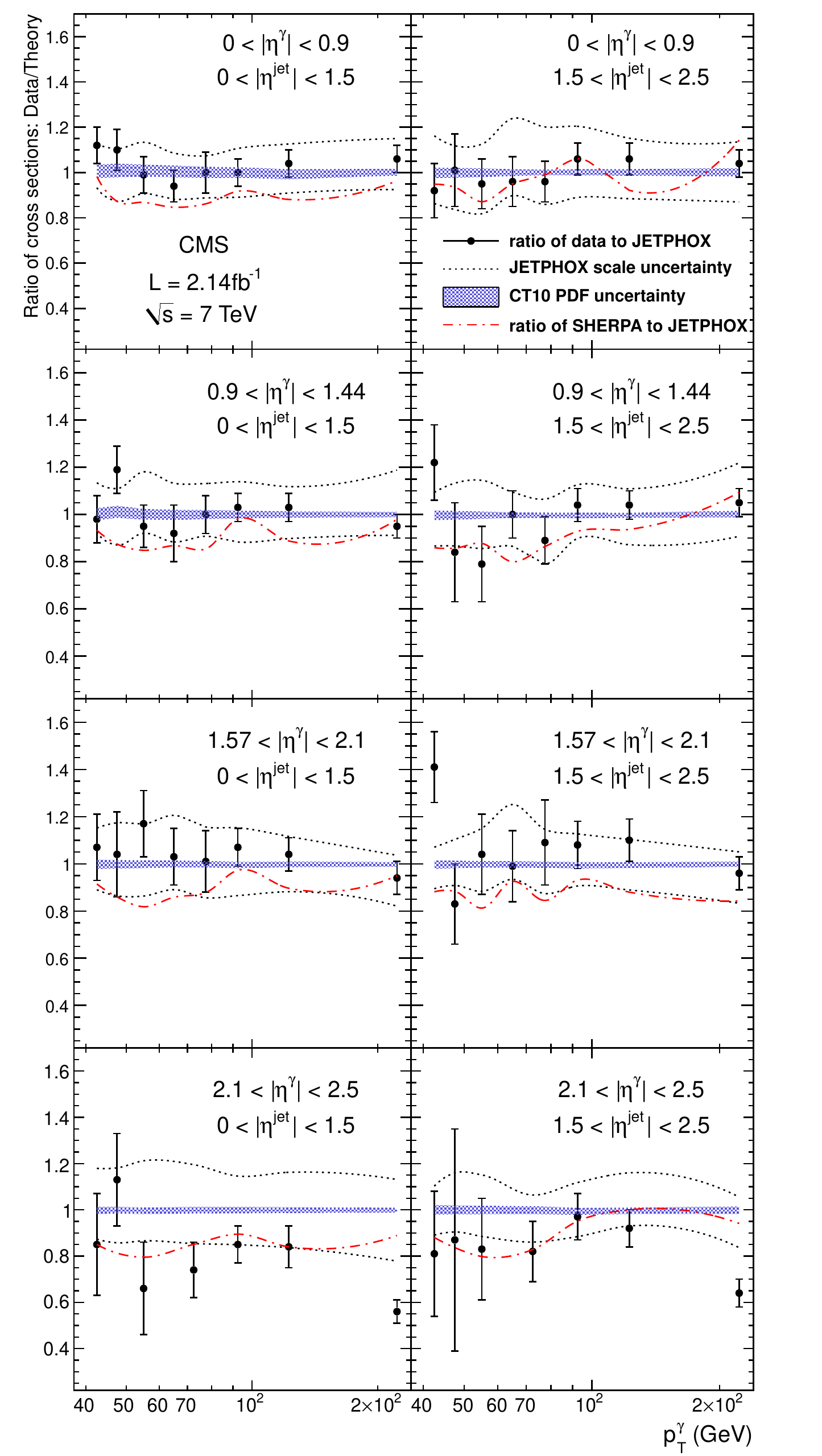}
\caption{The ratios of the measured triple-differential cross sections
to the NLO QCD prediction using \JETPHOX with the CT10 PDF set and scales
$\mu_{R,F,f} = \frac{1}{2}\pTg$. The vertical lines on the points show the
statistical and systematic uncertainties added in quadrature. The two dotted
lines represent the effect of varying the theoretical scales as described in the text.
The shaded bands correspond to the CT10 PDF uncertainty. The dash-dotted lines
show the ratios of the \SHERPA predictions to \JETPHOX.}
\label{fig:crossSection_comparison}
\end{center}
\end{figure*}

The measured cross sections are compared to theoretical predictions based
on perturbative QCD using the leading order (LO) MC event generator
\SHERPA (v1.3.1)~\cite{SHERPA} and the full next-to-leading order (NLO)
calculation implemented in \JETPHOX (v1.2.2)~\cite{jetphox}.
The SHERPA MC program incorporates higher-order tree level matrix elements
(ME) and parton shower (PS) modeling using the ME-PS matching algorithm
described in Ref.~\cite{multijet}. A similar technique is
also applied to processes involving prompt photons~\cite{hardphoton},
combining the photon and QCD parton multiplicity tree-level
matrix elements with a QCD+QED parton shower using the formalism
given in Ref.~\cite{multijet}, thus treating photons and jets
on an equal footing~\cite{hardphoton}.
This treatment also includes
contributions from the photon fragmentation component, permitting a direct
comparison with experimental measurements.  The predictions from \SHERPA
agree well with earlier photon measurements from the
Tevatron~\cite{Abazov:2008er}. The photon+jets final states are generated
with up to three additional jets using \SHERPA and the CTEQ6~\cite{cteq6}
parton distribution functions (PDFs).  Calculations are performed using
default choices for renormalization ($\mu_{R}$) and factorization ($\mu_{F}$)
scales equal to $\pTg$.
The \JETPHOX calculation at NLO in perturbative QCD includes
a model of fragmentation functions of partons to photons~\cite{BFG} and uses
the CT10~\cite{Lai:2010vv} NLO PDFs with
$\mu_{R}=\mu_{F}=\mu_{f}=\pTg/2$, where $\mu_{f}$ defines the fragmentation
scale.  To model the effect of experimental selection requirements
for these processes, the energy
around the photon within the $R<0.4$ cone is required to be less than 5\GeV.
The effect due to the choice of theory scales
is obtained by independently varying $\mu_{R}, \mu_{F}, \mu_{f}$ by the factors
0.5 and 2.0.  The uncertainty in the predictions due to the choice of PDF is
determined from the 40 (52) component error sets of CTEQ6M (CT10) and
evaluated using the master equations as given by the
`modified tolerance method' recommended in Ref.~\cite{lhapdfmethod}.
The effects of contributions from the
parton-to-hadron fragmentation and the underlying event
are examined by comparing cross sections determined using our default tune
in \PYTHIA at hadron level with and without multiple parton
interactions (MPI) and hadronization processes included.
We find these contributions to produce small fluctuations
around the parton-level cross section with little dependence on
kinematic variables and conclude that an uncertainty
of 1\% added to the \JETPHOX predictions in each (\etag, \etajet)
region covers their effects.
Figure~\ref{fig:crossSection_comparison} shows the ratios of the measured
triple-differential cross section to theoretical predictions.  The
determination of the photon signal purity contributes the main systematic
uncertainty affecting this measurement.
The central values of the cross section, the statistical uncertainty,
and the total systematic uncertainty are summarized in
Tables~\ref{tab:crossSection_pho_barrel} and~\ref{tab:crossSection_pho_endcap}.
The predictions from \SHERPA and \JETPHOX are consistent with data, except
for cases of photons measured in the largest $\eta$ and \pt\ regions.

\begin{figure*}[!hbt]
\begin{center}
\includegraphics[width=0.8\textwidth]{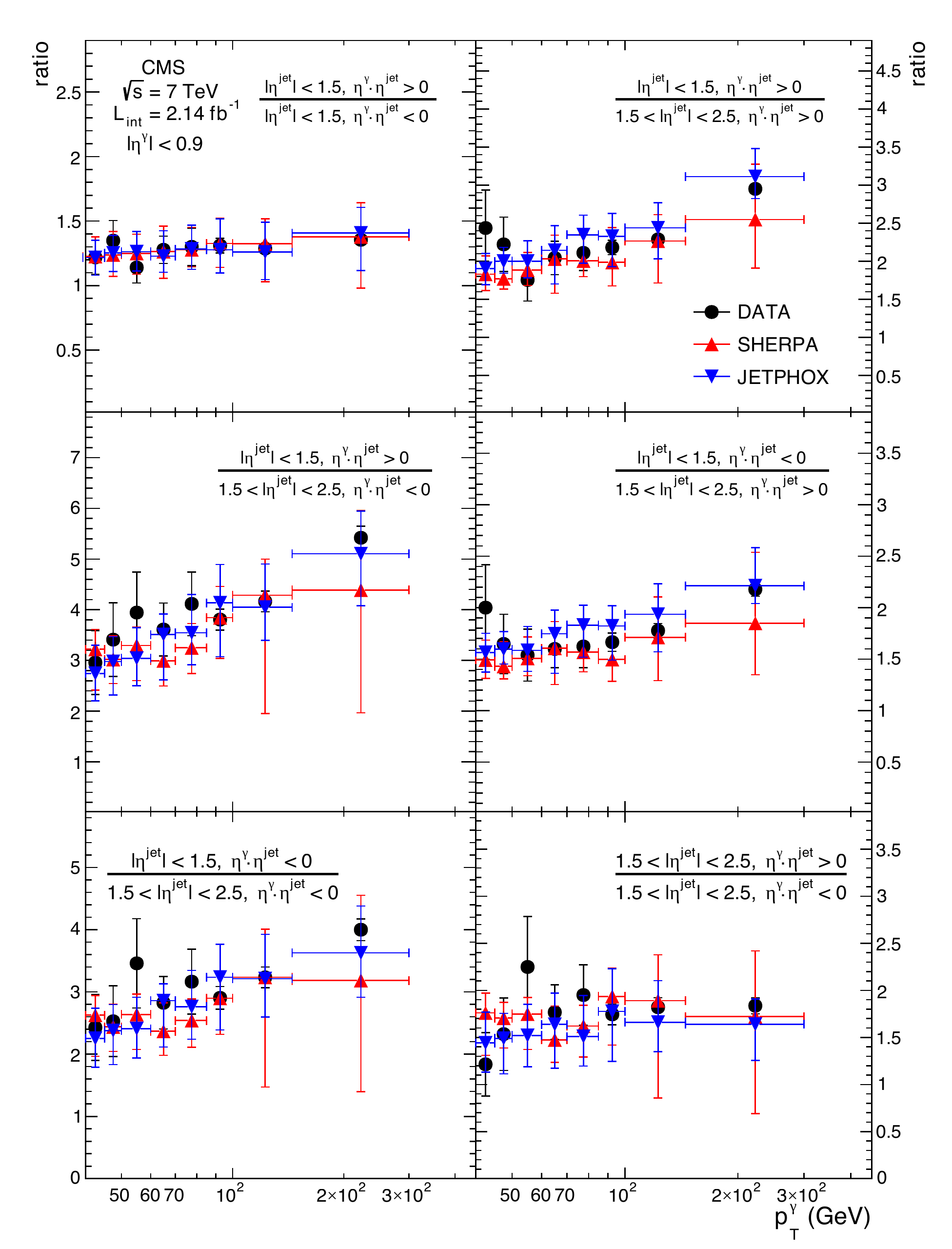}
\caption{Ratios of the triple-differential cross sections for the various jet
orientations with respect to the photon. The error bars on the theoretical
predictions correspond to statistical, scale and PDF uncetainties.}
  \label{fig:ratios}
  \end{center}
\end{figure*}

Figure~\ref{fig:ratios} shows the ratios of cross sections with different
angular orientations between the photon and the leading jet.  An earlier
study performed by the \DZERO experiment at the Tevatron~\cite{Abazov:2008er}
restricted the photon to $\abs{\etag}<1.0$, while allowing the jet to be either
in the central ($\abs{\etajet}<0.8$) or forward
($1.5<\abs{\etajet}<2.5$) region. In this study,
we consider $\abs{\etag} < 0.9$ and $\abs{\etajet}< 1.5$ or $1.5<\abs{\etajet}< 2.5$.
The advantage of measuring the ratios of cross sections is that
uncertainties in the integrated luminosity and reconstruction efficiencies largely cancel.

In conclusion, events with at least one photon and one jet have been studied
with a data sample corresponding to an integrated luminosity of
2.14\fbinv collected in proton-proton collisions at $\sqrt{s}=7$\TeV.
The cross section is measured as a function of the transverse momentum of the
photon for various configurations of the leading photon and the leading
jet.  These measurements are used
to determine eight ratios of the triple-differential cross section
\tdXS. They provide measures of the
relative cross sections for photon+jets production in
different pseudorapidity regions and, thus, over a wide range of parton
momentum fraction.  Comparisons of the data to
theoretical predictions from  \SHERPA  and \JETPHOX are also presented.
Although predictions from \SHERPA are observed to be lower than
those from \JETPHOX, the measured cross sections are found to be consistent
with both MC predictions within systematic uncertainties over most of the
measured kinematic regions.  The NLO predictions in QCD and
tree-level predictions of \SHERPA\ both fail to describe the data for
photons in the highest $\eta$ and \pt regions within expected variances
of either theoretical scale or parton distribution functions.

We congratulate our colleagues in the CERN accelerator departments
for the excellent performance of the LHC and thank the technical and
administrative staffs at CERN and at other CMS institutes for their
contributions to the success of the CMS effort. In addition, we gratefully
acknowledge the computing centers and personnel of the Worldwide LHC
Computing Grid for delivering so effectively the computing infrastructure
essential to our analyses. Finally, we acknowledge the enduring support for
the construction and operation of the LHC and the CMS detector provided
by the following funding agencies: BMWF and FWF (Austria);
FNRS and FWO (Belgium); CNPq, CAPES, FAPERJ, and FAPESP (Brazil);
MEYS (Bulgaria); CERN; CAS, MoST, and NSFC (China); COLCIENCIAS (Colombia);
MSES (Croatia); RPF (Cyprus); MoER, SF0690030s09 and ERDF (Estonia);
Academy of Finland, MEC, and HIP (Finland); CEA and CNRS/IN2P3 (France);
BMBF, DFG, and HGF (Germany); GSRT (Greece); OTKA and NKTH (Hungary);
DAE and DST (India); IPM (Iran); SFI (Ireland); INFN (Italy);
NRF and WCU (Republic of Korea); LAS (Lithuania);
CINVESTAV, CONACYT, SEP, and UASLP-FAI (Mexico); MSI (New Zealand);
PAEC (Pakistan); MSHE and NSC (Poland); FCT (Portugal);
JINR (Armenia, Belarus, Georgia, Ukraine, Uzbekistan);
MON, RosAtom, RAS and RFBR (Russia); MSTD (Serbia); SEIDI and CPAN (Spain);
Swiss Funding Agencies (Switzerland); NSC (Taipei);
ThEPCenter, IPST and NSTDA (Thailand); TUBITAK and TAEK (Turkey);
NASU (Ukraine); STFC (United Kingdom); DOE and NSF (USA).

\begin{table}[!ht]
\centering
\caption{The triple-differential cross sections \tdXS
for photons located in the central region with
statistical and systematic uncertainties, compared to predictions from
\JETPHOX and \SHERPA. A 2.2\% luminosity uncertainty is included in the
systematic uncertainty~\cite{CMS-PAS-SMP-12-008}.
The final two columns show the ratio of CMS data to \JETPHOX (D/J) and
\SHERPA (D/S), respectively.}
    \label{tab:crossSection_pho_barrel}
    {\small
    \begin{tabular}{|c|ccc|cc|}
    \hline
    \multicolumn{6}{|c|}{$\abs{\etag} < 0.9$ and $\abs{\etajet} < 1.5$} \\
    \hline
      $\pTg$ & \multicolumn{3}{c}{Cross section (pb/\GeVns{})}     & \multicolumn{2}{c|}{Ratio}\\
      \cline{2-6}
 (\GeVns{})    &          DATA                 & \JETPHOX & \SHERPA   &      D/J    &  D/S \\
\hline
40--45   & 27.9$\pm$1.0$\pm$1.8          &  24.9   &  24.5    &  1.12$\pm$0.08  &    1.14$\pm$0.08 \\
45--50   & 20.1$\pm$1.0$\pm$1.2          &  18.3   &  16.0    &  1.10$\pm$0.09  &    1.26$\pm$0.10 \\
50--60   & 10.70$\pm$0.40$\pm$0.77       &  10.8   &  9.41    &  0.99$\pm$0.08  &    1.14$\pm$0.09 \\
60--70   & 5.22$\pm$0.16$\pm$0.36        &  5.53   &  4.71    &  0.94$\pm$0.07  &    1.11$\pm$0.08 \\
70--85   & 2.62$\pm$0.09$\pm$0.21        &  2.61   &  2.26    &  1.00$\pm$0.09  &    1.16$\pm$0.10 \\
85--100  & 1.14$\pm$0.01$\pm$0.07        &  1.14   &  1.04    &  1.00$\pm$0.06  &    1.09$\pm$0.06 \\
100--145 & 0.358$\pm$0.003$\pm$0.020     &  0.344  &  0.303   &  1.04$\pm$0.06  &    1.18$\pm$0.07 \\
145--300 & 0.0320$\pm$0.0002$\pm$0.0018  &  0.0302 &  0.0290  &  1.06$\pm$0.06  &    1.10$\pm$0.06 \\
\hline
    \hline
    \multicolumn{6}{|c|}{$\abs{\etag} < 0.9$ and $1.5 < \abs{\etajet} < 2.5$}\\
    \hline
      $\pTg$ & \multicolumn{3}{c}{Cross section (pb/\GeVns{})} & \multicolumn{2}{c|}{Ratio}\\
      \cline{2-6}
(\GeVns{})     &         DATA                 & \JETPHOX & \SHERPA  &       D/J       & D/S \\
\hline
40--45   & 11.2$\pm$1.0$\pm$1.1         &  12.2   &  11.6   &  0.92$\pm$0.12  & 0.97$\pm$0.13 \\
45--50   & 8.59$\pm$0.82$\pm$1.05       &  8.52   &  7.94   &  1.01$\pm$0.16  & 1.08$\pm$0.17 \\
50--60   & 4.76$\pm$0.36$\pm$0.44       &  5.02   &  4.36   &  0.95$\pm$0.11  & 1.09$\pm$0.13 \\
60--70   & 2.19$\pm$0.14$\pm$0.21       &  2.29   &  2.17   &  0.96$\pm$0.11  & 1.01$\pm$0.11 \\
70--85   & 0.998$\pm$0.061$\pm$0.075    &  1.04   &  1.02   &  0.96$\pm$0.09  & 0.97$\pm$0.09 \\
85--100  & 0.454$\pm$0.009$\pm$0.028    &  0.429  &  0.455  &  1.06$\pm$0.07  & 1.00$\pm$0.06 \\
100--145 & 0.134$\pm$0.002$\pm$0.008    &  0.126  &  0.116  &  1.06$\pm$0.07  & 1.15$\pm$0.07 \\
145--300 & 0.0095$\pm$0.0001$\pm$0.0006 &  0.0091 &  0.0104 &  1.04$\pm$0.06  & 0.91$\pm$0.06 \\
\hline
  \hline
  \multicolumn{6}{|c|}{$0.9 < \abs{\etag} < \EtaBmax$ and $\abs{\etajet} < 1.5$}\\
  \hline
        $\pTg$ & \multicolumn{3}{c}{Cross section (pb/\GeVns{})} & \multicolumn{2}{c|}{Ratio}\\
        \cline{2-6}
(\GeVns{}) &              DATA                 & \JETPHOX & \SHERPA  &       D/J       & D/S \\
\hline
40--45   &  22.4$\pm$1.4$\pm$1.9         &  22.8   &  21.3   &  0.98$\pm$0.10  &  1.05$\pm$0.11\\
45--50   &  19.6$\pm$1.0$\pm$1.3         &  16.4   &  14.4   &  1.19$\pm$0.10  &  1.36$\pm$0.11\\
50--60   &  9.32$\pm$0.50$\pm$0.77       &  9.82   &  8.32   &  0.95$\pm$0.09  &  1.12$\pm$0.11\\
60--70   &  4.57$\pm$0.20$\pm$0.58       &  4.99   &  4.32   &  0.92$\pm$0.12  &  1.06$\pm$0.14\\
70--85   &  2.32$\pm$0.10$\pm$0.16       &  2.33   &  1.99   &  1.00$\pm$0.08  &  1.17$\pm$0.10\\
85--100  &  1.06$\pm$0.01$\pm$0.06       &  1.03   &  1.01   &  1.03$\pm$0.06  &  1.05$\pm$0.06\\
100--145 &  0.331$\pm$0.004$\pm$0.019    &  0.322  &  0.285  &  1.03$\pm$0.06  &  1.16$\pm$0.07\\
145--300 &  0.0283$\pm$0.0003$\pm$0.0016 &  0.0298 &  0.0291 &  0.95$\pm$0.05  &  0.97$\pm$0.06\\
\hline
  \hline
  \multicolumn{6}{|c|}{$0.9 < \abs{\etag} < \EtaBmax$ and $1.5 < \abs{\etajet} < 2.5$}  \\
	\hline
        $\pTg$ & \multicolumn{3}{c}{Cross section (pb/\GeVns{})} & \multicolumn{2}{c|}{Ratio}\\
        \cline{2-6}
(\GeVns{})     &          DATA                 & \JETPHOX & \SHERPA  &       D/J       & D/S\\
\hline
40--45   &  17.3$\pm$1.3$\pm$1.8         &  14.1   &  12.2   &  1.22$\pm$0.16  & 1.42$\pm$0.18\\
45--50   &  8.1$\pm$1.5$\pm$1.4          &  9.62   &  8.23   &  0.84$\pm$0.21  & 0.98$\pm$0.25\\
50--60   &  4.54$\pm$0.61$\pm$0.66       &  5.77   &  5.05   &  0.79$\pm$0.16  & 0.90$\pm$0.18\\
60--70   &  2.83$\pm$0.18$\pm$0.23       &  2.82   &  2.27   &  1.00$\pm$0.10  & 1.25$\pm$0.13\\
70--85   &  1.18$\pm$0.09$\pm$0.09       &  1.33   &  1.15   &  0.89$\pm$0.10  & 1.03$\pm$0.11\\
85--100  &  0.563$\pm$0.013$\pm$0.035    &  0.541  &  0.503  &  1.04$\pm$0.07  & 1.12$\pm$0.07\\
100--145 &  0.167$\pm$0.003$\pm$0.010    &  0.161  &  0.151  &  1.04$\pm$0.06  & 1.11$\pm$0.07\\
145--300 &  0.0121$\pm$0.0002$\pm$0.0007 &  0.0115 &  0.0127 &  1.05$\pm$0.06  & 0.96$\pm$0.06\\
\hline
\end{tabular}
    }
\end{table}

\begin{table}[!ht]	
\centering
\caption{The triple-differential cross sections
$d^3\sigma/(d\pTg d\etag d\etajet)$ for photons located in forward
region with statistical and systematic uncertainties, compared to
predictions from \JETPHOX and \SHERPA.
A 2.2\% luminosity uncertainty is included in the systematic uncertainty.
The final two columns show the ratio of CMS data to \JETPHOX (D/J) and \SHERPA (D/S),
respectively.}
    \label{tab:crossSection_pho_endcap}
    {\small
      \begin{tabular}{|c|ccc|cc|}
      \hline
      \multicolumn{6}{|c|}{$\EtaEmin < \abs{\etag} < 2.1$ and $\abs{\etajet} < 1.5$}  \\
	\hline
        $\pTg$ & \multicolumn{3}{c}{Cross section (pb/\GeVns{})} & \multicolumn{2}{c|}{Ratio}\\
        \cline{2-6}
(\GeVns{}) &              DATA                 & \JETPHOX & \SHERPA  &    D/J     & D/S\\
\hline
40--45   &  21.2$\pm$2.0$\pm$1.9         &  19.8   &  18.1   &  1.07$\pm$0.14  &  1.17$\pm$0.15\\
45--50   &  14.6$\pm$1.4$\pm$2.0         &  14.0   &  12.1   &  1.04$\pm$0.18  &  1.21$\pm$0.20\\
50--60   &  9.82$\pm$0.67$\pm$0.96       &  8.38   &  6.89   &  1.17$\pm$0.14  &  1.43$\pm$0.17\\
60--70   &  4.23$\pm$0.26$\pm$0.41       &  4.10   &  3.51   &  1.03$\pm$0.12  &  1.20$\pm$0.14\\
70--85   &  2.04$\pm$0.11$\pm$0.24       &  2.02   &  1.77   &  1.01$\pm$0.13  &  1.15$\pm$0.15\\
85--100  &  0.928$\pm$0.019$\pm$0.065    &  0.868  &  0.842  &  1.07$\pm$0.08  &  1.10$\pm$0.08\\
100--145 &  0.276$\pm$0.005$\pm$0.019    &  0.267  &  0.239  &  1.04$\pm$0.07  &  1.16$\pm$0.08\\
145--300 &  0.0221$\pm$0.0003$\pm$0.0017 &  0.0236 &  0.0223 &  0.94$\pm$0.07  &  0.99$\pm$0.08\\
\hline
        \hline
        \multicolumn{6}{|c|}{$\EtaEmin < \abs{\etag} < 2.1$ and $1.5 < \abs{\etajet} < 2.5$}  \\
	\hline
        $\pTg$ & \multicolumn{3}{c}{Cross section (pb/\GeVns{})} & \multicolumn{2}{c|}{Ratio}\\
        \cline{2-6}
(GeV)     &          DATA                 & \JETPHOX & \SHERPA  &        D/J      & D/S\\
\hline
40--45   &  22.3$\pm$1.4$\pm$1.9         &  15.8   &  14.0   &  1.41$\pm$0.15  & 1.60$\pm$0.17\\
45--50   &  9.1$\pm$1.4$\pm$1.1          &  10.9   &  9.66   &  0.83$\pm$0.17  & 0.94$\pm$0.19\\
50--60   &  6.92$\pm$0.68$\pm$0.86       &  6.65   &  5.39   &  1.04$\pm$0.17  & 1.28$\pm$0.20\\
60--70   &  3.13$\pm$0.21$\pm$0.43       &  3.15   &  2.92   &  0.99$\pm$0.15  & 1.07$\pm$0.16\\
70--85   &  1.63$\pm$0.11$\pm$0.25       &  1.50   &  1.26   &  1.09$\pm$0.18  & 1.29$\pm$0.22\\
85--100  &  0.694$\pm$0.017$\pm$0.059    &  0.643  &  0.596  &  1.08$\pm$0.10  & 1.16$\pm$0.10\\
100--145 &  0.202$\pm$0.004$\pm$0.016    &  0.183  &  0.162  &  1.10$\pm$0.09  & 1.25$\pm$0.10\\
145--300 &  0.0129$\pm$0.0002$\pm$0.0009 &  0.0135 &  0.0113 &  0.96$\pm$0.07  & 1.14$\pm$0.08\\
\hline
        \hline
        \multicolumn{6}{|c|}{$2.1 < \abs{\etag} < 2.5$ and $\abs{\etajet} < 1.5$}  \\
	\hline
        $\pTg$ & \multicolumn{3}{c}{Cross section (pb/\GeVns{})} & \multicolumn{2}{c|}{Ratio}\\
        \cline{2-6}
(GeV)     &         DATA                 & \JETPHOX & \SHERPA  & D/J      & D/S\\
\hline
40--45   & 14.5$\pm$3.4$\pm$1.6         &  17.1   &  14.5   &  0.85$\pm$0.22  & 1.00$\pm$0.26\\
45--50   & 13.6$\pm$2.0$\pm$1.4         &  12.0   &  9.77   &  1.13$\pm$0.20  & 1.39$\pm$0.25\\
50--60   & 4.72$\pm$0.76$\pm$1.2       &  7.17   &  5.71   &  0.66$\pm$0.20  & 0.83$\pm$0.25\\
60--85   & 1.78$\pm$0.16$\pm$0.25       &  2.42   &  2.05   &  0.74$\pm$0.12  & 0.87$\pm$0.14\\
85--100  & 0.607$\pm$0.031$\pm$0.048    &  0.713  &  0.641  &  0.85$\pm$0.08  & 0.95$\pm$0.09\\
100--145 & 0.174$\pm$0.008$\pm$0.016    &  0.206  &  0.174  &  0.84$\pm$0.09  & 1.00$\pm$0.10\\
145--300 & 0.0082$\pm$0.0004$\pm$0.0007 &  0.0145 &  0.0129 &  0.56$\pm$0.05  & 0.63$\pm$0.06\\
\hline
        \hline
        \multicolumn{6}{|c|}{$2.1 < \abs{\etag} < 2.5$ and $1.5 < \abs{\etajet} < 2.5$}  \\
	\hline
        $\pTg$ & \multicolumn{3}{c}{Cross section (pb/\GeVns{})} & \multicolumn{2}{c|}{Ratio}\\
        \cline{2-6}
(\GeVns{})    &         DATA                 & \JETPHOX & \SHERPA  &       D/J       &  D/S\\
\hline
40--45   & 13.2$\pm$4.2$\pm$1.4         &  16.2   &  14.4   &  0.81$\pm$0.27  &  0.92$\pm$0.31\\
45--50   & 9.9$\pm$4.0$\pm$3.7          &  11.4   &  9.51   &  0.87$\pm$0.48  &  1.04$\pm$0.57\\
50--60   & 5.6$\pm$1.0$\pm$1.0          &  6.75   &  5.36   &  0.83$\pm$0.22  &  1.04$\pm$0.27\\
60--85   & 1.87$\pm$0.18$\pm$0.23       &  2.29   &  1.88   &  0.82$\pm$0.13  &  0.99$\pm$0.16\\
85--100  & 0.607$\pm$0.029$\pm$0.054    &  0.628  &  0.593  &  0.97$\pm$0.10  &  1.02$\pm$0.10\\
100--145 & 0.148$\pm$0.006$\pm$0.012    &  0.160  &  0.161  &  0.92$\pm$0.08  &  0.92$\pm$0.08\\
145--300 & 0.0060$\pm$0.0003$\pm$0.0005 &  0.0094 &  0.0088 &  0.64$\pm$0.06  &  0.68$\pm$0.06\\
\hline
\end{tabular}
    }
\end{table}

\clearpage

\bibliography{auto_generated}   
\clearpage

\cleardoublepage \appendix\section{The CMS Collaboration \label{app:collab}}\begin{sloppypar}\hyphenpenalty=5000\widowpenalty=500\clubpenalty=5000\input{QCD-11-005-authorlist.tex}\end{sloppypar}
\end{document}

%% file: QCD-11-005-authorlist.tex
\textbf{Yerevan Physics Institute,  Yerevan,  Armenia}\\*[0pt]
S.~Chatrchyan, V.~Khachatryan, A.M.~Sirunyan, A.~Tumasyan
\vskip\cmsinstskip
\textbf{Institut f\"{u}r Hochenergiephysik der OeAW,  Wien,  Austria}\\*[0pt]
W.~Adam, T.~Bergauer, M.~Dragicevic, J.~Er\"{o}, C.~Fabjan\cmsAuthorMark{1}, M.~Friedl, R.~Fr\"{u}hwirth\cmsAuthorMark{1}, V.M.~Ghete, N.~H\"{o}rmann, J.~Hrubec, M.~Jeitler\cmsAuthorMark{1}, W.~Kiesenhofer, V.~Kn\"{u}nz, M.~Krammer\cmsAuthorMark{1}, I.~Kr\"{a}tschmer, D.~Liko, I.~Mikulec, D.~Rabady\cmsAuthorMark{2}, B.~Rahbaran, C.~Rohringer, H.~Rohringer, R.~Sch\"{o}fbeck, J.~Strauss, A.~Taurok, W.~Treberer-Treberspurg, W.~Waltenberger, C.-E.~Wulz\cmsAuthorMark{1}
% ...
\vskip\cmsinstskip
% ...
1:~~Also at Vienna University of Technology, Vienna, Austria\\
2:~~Also at CERN, European Organization for Nuclear Research, Geneva, Switzerland\\
% ...
